\documentclass{article}

\usepackage[final,nonatbib]{neurips_2024}

\usepackage[utf8]{inputenc} 
\usepackage[T1]{fontenc}    
\usepackage[hidelinks]{hyperref}       
\usepackage{url}            
\usepackage{booktabs}       
\usepackage{amsmath} 
\usepackage{bm} 
\usepackage{siunitx} 
\usepackage{amsfonts}       
\usepackage{nicefrac}       
\usepackage{microtype}      
\usepackage{xcolor}         
\usepackage{graphicx}
\usepackage{longtable}
\usepackage{enumitem}       
\usepackage[backend=biber,style=ieee,citestyle=numeric-comp]{biblatex}
\usepackage{pythonhighlight}
\usepackage{listings}
\usepackage{color}
\definecolor{gray}{rgb}{0.5,0.5,0.5}
\newcommand{\ie}{\textit{i.e.~}}
\newcommand{\cf}{\textit{cf.~}}
\newcommand{\etal}{\textit{et al.}}
\newcommand{\smolnet}{SMolNet}

\title{Automatic solid form classification in pharmaceutical drug development}

\author{
  Julius Lange\textnormal{\textsuperscript{1}}  \And  Leonid Komissarov\textnormal{\textsuperscript{1}} \And  Rene Lang\textnormal{\textsuperscript{1}}  \AND  Dennis Dimo Enkelmann\textnormal{\textsuperscript{2}} \And Andrea Anelli\textnormal{\textsuperscript{1}}\\ \AND
  \\
  \textnormal{\textsuperscript{1}} Roche Pharmaceutical Research and Early Development\\
  \textnormal{\textsuperscript{2}} Pharma Technical Development\\ 
  F. Hoffmann-La Roche Ltd.\\
  Basel, Switzerland\\
  \texttt{andrea.anelli@roche.com}
}

\begin{document}

\maketitle

\begin{abstract}
In materials and pharmaceutical development, rapidly and accurately determining the similarity between X-ray powder diffraction (XRPD) measurements is crucial for efficient solid form screening and analysis.
We present SMolNet, a classifier based on a Siamese network architecture, designed to automate the comparison of XRPD patterns.
Our results show that training SMolNet on loss functions from the self-supervised learning domain yields a substantial boost in performance with respect to class separability and precision,
specifically when classifying phases of previously unseen compounds.
The application of SMolNet demonstrates significant improvements in screening efficiency across multiple active pharmaceutical ingredients, providing a powerful tool for scientists to discover and categorize measurements with reliable accuracy.
\end{abstract}

\section{Introduction}\label{sec:intro}
\begin{figure}
    \centering
\includegraphics[width=1.0\textwidth, angle=0]{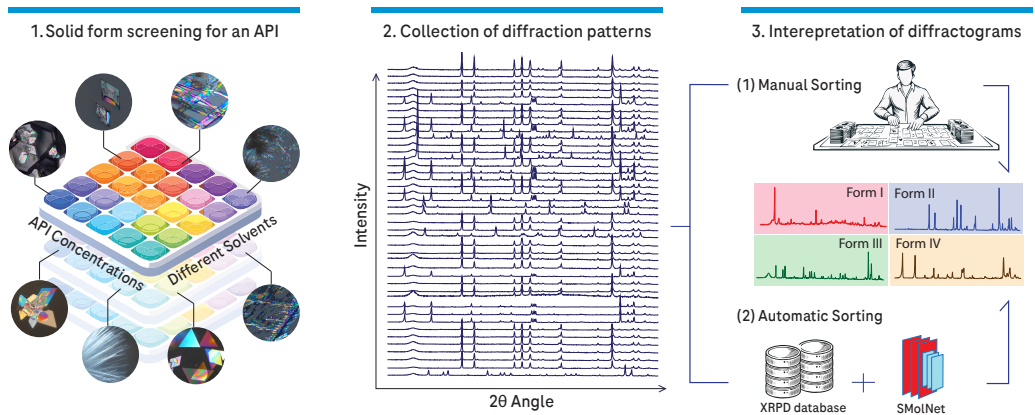}
   \caption{Example of a solid form screening project: (1) A candidate molecule is mixed in different solutions to explore its solid state landscape. Measurable crystalline residues that are formed are investigated using XRPD - creating a large number of measurements of potentially different materials (2). To identify the different forms obtained, typically a manual sorting is employed (3-1). Alternatively, we propose to leverage previous projects' data to train SMolNet, which can be used to perform the same sorting task at higher throughput (3-2).}
   \label{fig:workflow}
\end{figure}

Advancements in artificial intelligence have revolutionized various fields by automating complex, experience-driven processes.
In pharmaceutical solid-state development, similar to materials science, there is a growing need to automate the analysis of X-ray powder diffraction (XRPD) patterns \cite{leeman2024challenges}, which are critical for identifying and characterizing the solid form landscape of active pharmaceutical ingredients (APIs) \cite{Brittain, Byrn, Hilfiker, Bernstein,BRAUN20092010, Braun2017Complexity}.
Here, multiple patterns from crystallization experiments of the same API are compared and assigned to a set of reference diffractograms, each representing a distinct solid form of the compound of interest, as depicted in Figure~\ref{fig:workflow}.
Traditional methods for XRPD pattern comparison often require expert knowledge and manual inspection, making the process time-consuming and susceptible to human error \cite{klug1954xray, guinier1994xray}.
Moreover, manual interpretation becomes impractical when dealing with hundreds or thousands of samples generated from high throughput crystallization experiments aimed at discovering new solid forms \cite{Brittain2, Reutzel-Edens:yc5024}.

Previous work to automate the classification of XRPD patterns includes measures operating directly on the signals \cite{degelder, karfunkel}, CNN classifiers \cite{cnn-xrd-example, VGG16} and Siamese networks \cite{jan1}.
A key limitation in many of these works consists in framing this exercise into a classification task where the reference forms are known \textit{a priori} \cite{jan2,ceder}.
In practice however, a new unseen form can be observed at every measurement during experimental screening.
Further, most of the research in this field focuses on highly crystalline inorganic materials that typically exhibit simpler and sharper diffraction peaks due to high symmetry and rigid crystal structures.
Here the generation of synthetic XRPD patterns is a common technique for enrichment of training data \cite{cnn-xrd-example, gen-xrpd1, gen-xrpd2, gen-xrpd3}.
Small molecule APIs however, typically produce more complex XRPD patterns with broader peaks and additional artifacts, stemming from their intricate molecular structures, lower symmetry, and variations such as amorphous backgrounds, height preparation shifts, preferential orientation, and particle size variance.
These complexities are especially evident in real-life data obtained from solid form screening routines of organic APIs. Such variations pose significant challenges for automated pattern recognition and classification, as methods developed for inorganic materials may not transfer directly to the organic domain. \cite{warren1990x, ermrich2013xrd}. 

To address these challenges, we introduce SMolNet (\textbf{S}olid-form \textbf{Mol}ecules \textbf{Net}work), a novel deep learning framework specifically designed for the pairwise classification and identification of XRPD patterns of organic crystalline materials.
Our approach leverages Siamese networks to capture subtle similarities and differences between complex diffraction patterns, even with limited training data.
By training directly on experimental XRPD data from organic crystals, our model bypasses the need of synthetic data augmentation and demonstrates reliable generalization to new, unseen patterns and chemical spaces.
Our contributions are as follows:
\begin{itemize}[itemsep=-1pt,leftmargin=*]
    \item We propose a Siamese network architecture, trained on a modified SigLIP loss \cite{SigLIP} for the first time to handle organic XRPD patterns in a zero shot learning setting.
    \item Robust training and testing on experimental data using a leave-two-compounds-out routine (L2CO)
    \item Improved classification performance:
    We demonstrate significant improvements over non data-driven methods and previous architectures in discriminating across different forms.
\end{itemize}
By bridging the gap between machine learning and materials science, our work advances the application of AI in pharmaceutical solid-state development. This not only accelerates the solid-form screening process but also paves the way for more efficient and accurate materials discovery.

\section{Methods}\label{sec:method}

\subsection{Data}\label{sec:data}
The dataset constructed for this study is proprietary and contains $3750$ experimental XRPD patterns measured across $16$ organic compounds that constitute drug candidates in a pharmaceutical research and development setting, for a total of $24$ different solid forms. Each pattern constitutes a $1$-d signal of $1950$ points, covering a scattering angle of $2\theta \in [3, 42)$. 
Additional experimental details are provided in Appendix \ref{sec:appendix:experimental}.
For each compound, patterns that belong to the same phase were manually labelled by experimentalists.
All pattern intensities were normalized to be within the $[0, 1]$ range.
We consider pairwise combinations of individual patterns, amounting to roughly \num{7e6} pairs and a respective label specifying whether the two patterns belong to the same phase (positive class) or not (negative class).
The ratio of positive to negative classes is $1:9.5$, creating a substantially imbalanced dataset.
We remark how this large curated database constitutes, to the best of our knowledge, a first example on the feasibility of training a foundational model for organic solid diffraction patterns recognition.

\subsection{Model architecture}\label{sec:model}
We aim to classify whether two diffraction patterns $\bm{x}_1$ and $\bm{x}_2$ belong to the same phase or not.
To that end we consider $\bm{z} = f(\bm{x}, \bm{\theta})$ --
a function that translates individually normalized input intensities of a pattern $\bm{x} \in \mathbb{R}^{1950}$ to a latent embedding $\bm{z} \in \mathbb{R}^n$,
where $f$ can additionally be parameterized by $\bm{\theta}$.
For two patterns representing the phase we would like the distance of their latent embeddings
$d(\bm{z}_1, \bm{z}_2)$
to be small and vice versa.
We find the best architecture for this task through hyperparameter optimization.
Our model constitutes four $1$-dimensional convolutional layers with a kernel size of $8$ to $128$,
each followed by a batch normalization, Mish activation \cite{mish} and dropout with a probability of $0.2$.
All convolutional outputs are concatenated and passed on to a multilayer perceptron, consisting of two dense layers with $2048$ hidden neurons.
The final output embedding is $\bm{z} \in \mathbb{R}^{128}$.
When trained on the Contrastive loss (\cf following section), the last layer embeddings are additionally  passed through a Sigmoid activation function.
A detailed diagram of the architecture is provided in Apppendix \ref{sec:appendix:architecture}.

\subsection{Training}\label{sec:training}
To simulate a prospective application setting as closely as possible while guaranteeing statistical significance of the results, a leave-two-compounds-out (L2CO) cross-validation technique is employed.
At each fold a unique combination of two out of the total of $16$ compounds are selected and all measurements belonging to one of the two compounds are used for testing, resulting in a total of $120$ folds (\cf Sec.\ \ref{sec:data}).
This ensures presence of at least two forms in the test set to have both positive and negative pairs.
Of the remaining $14$ compounds, $11$ are used for training and three, randomly chosen, for validation.
The model weights are optimized with the Adam \cite{adam} optimizer at an initial learning rate of \num{5e-3}.
Best model parameters are saved every time a new minimal loss is observed on the validation set.
The learning rate is reduced by a factor of $0.1$ if no improvement in the validation loss is observed after $15$ evaluations.
The model is trained for up to $20$ epochs.

We train and compare the SMolNet performance after training on one of three different loss functions, each previously applied to contrastive learning tasks.
The functions tested are the
Contrastive \cite{contrastive-loss1, contrastive-loss2,},
NT-XEnt/InfoNCE \cite{InfoNCE1, InfoNCE2, INfoNCE3}
and SigLIP \cite{SigLIP} losses.
Considering our application we propose two modifications to the SigLIP loss.
First, rather than using a cosine similarity for training the embeddings, the Euclidean distance is used to generate the logit values.
This is motivated by the improved classification performance, as reported in Appendix \ref{sec:appendix:results}.
Second, since we are not dealing with a self-supervised task, we relax the criterion that all matching pairs are only
found along the diagonal of the pairwise matrix of labels.
We present a pseudo-implementation of the modified SigLIP algorithm in Appendix \ref{sec:appendix:siglip}.
When training with the Contrastive loss, weighted sampling is employed during the training to account for class imbalance.
Training times with the Contrastive loss are approximately $15$-$25$ mins for one fold on a NVIDIA A100 / LS40S GPU.
Significant speedup by a factor of $5$-$10$ is achieved when training with one of the other loss functions.
This is due to the explicit generation of input pairs when training with the Contrastive loss, effectively resulting in $\sim N^2$ data entries. 

To provide a lower bound for predictivity, we compare the performance of SMolNet to a naive baseline model
where the score is directly computed from the euclidean distance such that:
$d(\bm{z}_1, \bm{z}_2) = ||\bm{x}_1-\bm{x}_2||$.
We follow the same cross-validation procedure as above,
computing all pairwise euclidean distances on the raw $1$-d input signals in the training data, interpreting the results as classifier probabilities.
We then choose a threshold that maximizes the F1 score and apply it to the test set patterns in order to derive predicted binary labels. 
Additionally, we compare the performance of SMolNet to a previously published CNN architecture \cite{jan1, VGG16} and a similarity measure by de Gelder \etal ~\cite{degelder}.
Both methods were developed specifically for the classification of XRPD patterns.

\section{Results}\label{sec:results}
Average results of the L2CO cross-validation are presented in Table \ref{tab:results},
showing the area under the ROC curve (AUROC), area under the precision-recall curve (AUPRC), accuracy and F1 score.
Note that the AUPRC value associated with a random model is variable, equaling to the ratio of positive labels in the test set, $\text{AUPRC}_{\text{random}}=\nicefrac{N_\text{positive}}{N_\text{total}}$.
We thus report this metric as the delta above the baseline value: $\text{AUPRC}=\text{AUC}(R,P)-\text{AUPRC}_{\text{random}}$.

Our findings reveal that the naive and de Gelder models already perform well in separating positive from negative classes with an above-average precision.
Hard classifier metrics, \ie Accuracy and F1 score are best for these models, hinting that threshold finding might be easiest when using simple approaches that operate directly on the input patterns.  
Nevertheless, machine learning approaches clearly improve on both AUROC and AUPRC metrics.
This is particularly important when evaluating cases as ours, where a high class-imbalance (\cf Section \ref{sec:data}) can produce overly-optimistic accuracy and F1 scores.
We therefore argue that evaluating models on the soft classifier metrics is more meaningful if bias to a particular class should be avoided.
We observe marginal improvements in all metrics when comparing SMolNet to a previously published architecture \cite{jan1, VGG16} when using a Contrastive loss.
Notably, training on the NT-XEnt and SigLIP loss functions appears to be highly advantageous when high average precision and class separability are desired. 
Our best-performing model is SMolNet trained on the SigLIP loss, resulting
in average AUROC and AUPRC values of 0.98 and 0.65, respectively.
Conversely, applying the SigLIP loss to a previously published architecture \cite{jan1, VGG16} leads to further degradation of performance, as reported in Appendix \ref{sec:appendix:results}.
Although the de Gelder method produces the best F1 score, we argue that the increased AU\{ROC,PRC\}  scores of the machine learning models is particularly significant in typical human-in-the-loop scenarios where partial labeling is performed preliminarily on a subset of data points.
This allows for further threshold optimization on test data, making SMolNet especially valuable in real-world applications where efficiency and accuracy are paramount.
For additional experiments and relevant loss funciton parameters please refer to Appendix \ref{sec:appendix:results}.

\begin{table}[h!]
    \centering
    \caption{
        Average L2CO performance and their 95\% confidence interval for various tested approaches. 
        Showing the area under the ROC curve (AUROC),
        area under the precision-recall curve (AUPRC),
        Accuracy and F1 score.
        Best and second-best results in bold and underlined, respectively.
    }
    \label{tab:results}

    \begin{tabular}{llrrrr}
        \toprule
        Method & Loss function\hspace{2em} & AUROC & AUPRC & Acc & F1 \\
        \midrule
        Naive & & $0.88_{\pm.14}$ & $0.34_{\pm.26}$ & $\underline{0.89}_{\pm.14}$ & $\underline{0.90}_{\pm.13}$ \\
        
        De Gelder \etal\supercite{degelder} & & $0.92_{\pm.09}$ & $0.46_{\pm.21}$ & $\bm{0.92}_{\pm.10}$ & $\bm{0.93}_{\pm.11}$ \\
        
        Schützke \etal\supercite{jan1, VGG16} & Contrastive & $0.95_{\pm.14}$ & $0.41_{\pm.25}$ & $0.86_{\pm.20}$ & $0.87_{\pm.18}$ \\
        
        \smolnet & Contrastive & $0.97_{\pm.09}$ & $0.43_{\pm.24}$ & $0.87_{\pm.20}$ & $0.89_{\pm.17}$ \\
        
        \smolnet & NT-XEnt & $\underline{0.97}_{\pm.08}$ & $\underline{0.63}_{\pm.33}$ & $0.89_{\pm.20}$ & $0.79_{\pm.39}$ \\
        
        \smolnet & SigLIP & $\bm{0.98}_{\pm.09}$ & $\bm{0.65}_{\pm.31}$ & $0.89_{\pm.21}$ & $0.80_{\pm.38}$ \\
        \bottomrule
    \end{tabular}

\end{table}

\section{Conclusion \& Outlook}\label{sec:conclusion}
We have provided a model architecture and training setup with extensive proof of its performances across different pharmaceutical materials. The ability to reliably distinguish between similar and dissimilar patterns across different organic compounds and compositions constitutes a fundamental first step in the direction of tackling this problem with a foundational model approach. 
Our findings further report that the incorporation of contrastive loss functions into the model training procedure is highly beneficial.
While our findings show that SMolNet provides superior class separation based on soft metrics, we recognize that optimal threshold determination for classification remains a challenge. Threshold selection is strongly influenced by training procedures and is highly dependent on the test data, particularly in zero-shot learning scenarios where test compounds are not present in the training data. Addressing this issue requires further investigation into adaptive thresholding methods and calibration techniques to improve generalization to unseen compounds.
Having established a reliable architecture in complex pure phase projects, we plan in future to explore the integration of measurements from public databases, as well as synthetic data coming from crystal structure prediction simulations. Finally, by combining the multiple single phase data we aim to tackle the challenging task of mixture determination.

\begin{ack}
L. K. acknowledges funding by the Roche Postdoctoral Fellowship (RPF) programme. The authors thank Leoni Grossman, Andre Egli and Joost van den Ende for inspiring discussions.
\end{ack}

\section*{References}
\AtNextBibliography{\small}
\printbibliography[heading=none]

\newpage
\appendix

\section{SMolNet Architecture}\label{sec:appendix:architecture}
Architecture of SMolNet described in Sec.\ \ref{sec:method}.
\begin{figure}[h!]
    \centering
   \includegraphics[width=1.\textwidth, angle=0]{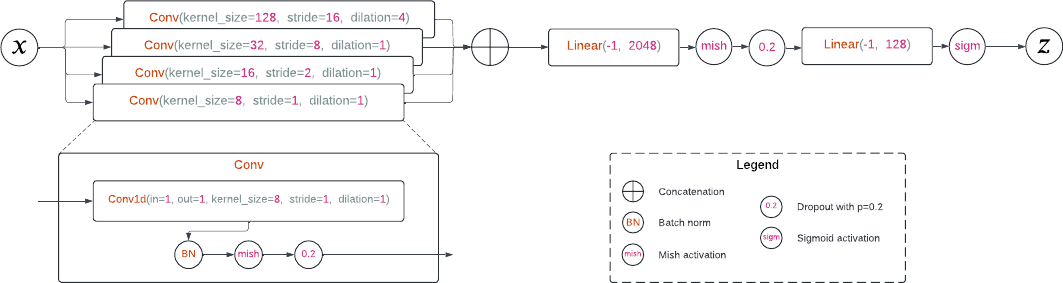}
   \caption{
   SMolNet architecture diagram with hyperparameters used in this study.
   }
   \label{fig:architecture}
\end{figure}

\section{Modified SigLIP Pseudo-Implementation}\label{sec:appendix:siglip}

\begin{python}
# n : batch size
# emb1 : model embeddings to be compared with emb2 [n, dim]
# emb2 : model embeddings to be compared with emb1 [n, dim]
# labels1 : labels of emb1 [n,1]
# labels2 : labels of emb2 [n,1]
# t_prime, b : learnable temperature and bias

t = exp(t_prime)
z1 = l2_normalize(emb1)
z2 = l2_normalize(emb2)
logits = l2_distance(z1, z2) * t + b # [n,n]

labels = int(labels1 == labels2.T) # [n,n], positive pairs: 1
labels[labels==0] = -1 # negative pairs: -1

loss = -sum(log_sigmoid(labels * logits)) / n

\end{python}

\section{Additional Experimental Results and Parameters}\label{sec:appendix:results}
\begin{table}[h!]
    \centering
    \caption{
        Average L2CO performance and their 95\% confidence interval for various tested approaches,
        and parameters used.
        Showing the area under the ROC curve (AUROC),
        area under the precision-recall curve (AUPRC),
        Accuracy and F1 score.
    }

    \resizebox{\textwidth}{!}{
    \begin{tabular}{lllrrrr}
        \toprule
        Method & Loss function\hspace{2em} & Parameters & AUROC & AUPRC & Acc & F1 \\
        \midrule
        Naive & &  & $0.88_{\pm.14}$ & $0.34_{\pm.26}$ & ${0.89}_{\pm.14}$ & ${0.90}_{\pm.13}$ \\
        
        De Gelder \etal\supercite{degelder} & & $l=0.005$ & $0.92_{\pm.09}$ & $0.46_{\pm.21}$ & ${0.92}_{\pm.10}$ & ${0.93}_{\pm.11}$ \\
        
        Schützke \etal\supercite{jan1, VGG16} & Contrastive & $m=8.0$ & $0.95_{\pm.14}$ & $0.41_{\pm.25}$ & $0.86_{\pm.20}$ & $0.87_{\pm.18}$ \\
        Schützke \etal                        & SigLIP & L2 distance & $0.69_{\pm.29}$ & $0.19_{\pm.40}$ & $0.39_{\pm.48}$ & $0.41_{\pm.44}$ \\
        
        \smolnet & Contrastive & $m=8.0$ & $0.97_{\pm.09}$ & $0.43_{\pm.24}$ & $0.87_{\pm.20}$ & $0.89_{\pm.17}$ \\
        \smolnet & NT-XEnt & L2 distance, $T=5.0$ & ${0.97}_{\pm.08}$ & ${0.63}_{\pm.33}$ & $0.89_{\pm.20}$ & $0.79_{\pm.39}$ \\
        \smolnet & SigLIP & L2 distance & ${0.98}_{\pm.09}$ & ${0.65}_{\pm.31}$ & $0.89_{\pm.21}$ & $0.80_{\pm.38}$ \\
        \smolnet & SigLIP & Cosine similarity & $0.94_{\pm.20}$ & $0.56_{\pm.43}$ & $0.83_{\pm.36}$ & $0.74_{\pm.45}$ \\
        \bottomrule
    \end{tabular}
    }
\end{table}

\section{Experimental Setup for XRPD measurements}\label{sec:appendix:experimental}
The dataset used in this study consists of X-ray diffraction (XRD) measurements collected from 2006 to the present, across various research and development projects, using STOE STADI P diffractometers. These systems were equipped with curved germanium (Ge(111)) monochromators and Cu K$\alpha$1 radiation sources ($\lambda$ = $1.54060$\AA). The instruments operated at a voltage of $40$ kV, with currents ranging from $40$ mA to $50$ mA, depending on the specific experimental setup.
Samples weighing $1$ to $5$ mg were placed in sample cells with aperture diameters between $3$ mm and $5$ mm and a sample depth of $0.45$ mm. For wet samples, Kapton film clips were used, while cellulose acetate film clips were employed for dry samples. Data were recorded in transmission mode, spanning an angular range of $3\deg$ to $42\deg$ in $2\theta$, using a moving Position Sensitive Detector (PSD) with a fixed omega angle.
Over the years, two detector systems were employed for data collection. Initially, the STOE Linear Position Sensitive Detector (PSD) was used, covering up to $6\deg$ in $2\theta$ per scan, allowing for rapid data acquisition. A step size of $0.5\deg$ in $2\theta$ was applied, with dwell times ranging from $5$ to $40$ seconds per step, depending on the measurement mode (standard or rapid).
Subsequent measurements were performed with Dectris MYTHEN K1 and MYTHEN K2 strip detectors, which offered a broader angular coverage of $12.5\deg$ in $2\theta$ and a high resolution of $0.01\deg$ in $2\theta$. For both MYTHEN detectors, the step size remained at $0.5\deg$ in $2\theta$, with dwell times ranging from $5$ to $20$ seconds per step, depending on whether standard or rapid measurements were taken.

\end{document}